\documentstyle[prl,aps,epsfig]{revtex}

\begin{document}

\draft

\title{Evidence for a Transition in the Pairing Symmetry of the
Electron-Doped Cuprates La$_{2-x}$Ce$_x$CuO$_{4-y}$ and
Pr$_{2-x}$Ce$_x$CuO$_{4-y}$}

\author{John A. Skinta, Mun-Seog Kim, and Thomas R. Lemberger}
\address{Department of Physics, Ohio State University, Columbus,
OH 43210-1106}
\author{T. Greibe and M. Naito}
\address{NTT Basic Research Laboratories, 3-1 Morinosato Wakamiya,
Atsugi-shi, Kanagawa 243, Japan}

\date{\today}
\maketitle

\begin{abstract}

We present measurements of the magnetic penetration depth,
$\lambda^{-2}(T)$, in Pr$_{2-x}$Ce$_{x}$CuO$_{4-y}$ and
La$_{2-x}$Ce$_{x}$CuO$_{4-y}$ films at three Ce doping levels,
$x$, near optimal. Optimal and overdoped films are qualitatively
and quantitatively different from underdoped films. For example,
$\lambda^{-2}(0)$ decreases rapidly with underdoping but is
roughly constant above optimal doping. Also, $\lambda^{-2}(T)$ at
low $T$ is exponential at optimal and overdoping but is quadratic
at underdoping. In light of other studies that suggest both
\textit{d}- and \textit{s}-wave pairing symmetry in nominally
optimally doped samples, our results are evidence for a transition
from \textit{d}- to \textit{s}-wave pairing near optimal doping.

\end{abstract}


A variety of experiments have demonstrated that hole-doped
cuprates possess a predominantly $d_{x^2-y^2}$ gap
\cite{harlingen,tsuei01}. In contrast, there is no consensus
concerning order parameter symmetry in electron-doped cuprates.
Phase-sensitive \cite{tsuei03}, angel resolved photoemission
spectroscopy \cite{armitage}, and some penetration depth
\cite{kokales01,prozorov} measurements on nominally identical
optimally doped Pr$_{2-x}$Ce$_{x}$CuO$_{4-y}$ (PCCO) and
Nd$_{2-x}$Ce$_{x}$CuO$_{4-y}$ (NCCO) samples suggest
\textit{d}-wave pairing. Other penetration depth measurements
\cite{alff01,skinta01} and the absence of a zero-bias conductance
peak in tunnelling measurements \cite{kashiwaya,alff02} indicate
\textit{s}-wave superconductivity.

To date, experimental studies of e-doped cuprates have
concentrated on optimally doped samples. To explore the pairing
symmetry controversy, we present measurements of $\lambda^{-2}(T)$
in PCCO and La$_{2-x}$Ce$_{x}$CuO$_{4-y}$ (LCCO) films at three
dopings, $x$, near optimal. The curvature near $T_c$ and the
low-temperature magnitude of the superfluid density, $n_s(T)
\propto \lambda^{-2}(T)$, depend strongly on doping. Furthermore,
the temperature dependence of $\lambda^{-2}(T)/\lambda^{-2}(0)$ at
low $T$ changes with doping: it is exponential at optimal and
overdoping, but quadratic at underdoping. These phenomena indicate
some sort of transition near optimal doping. Our results here are
consistent with the transition being from \textit{d}- to
\textit{s}-wave pairing. Contradictory e-doped pairing symmetry
results can thus be reconciled, if nominally optimally doped
samples that exhibit \textit{d}-wave properties are in reality
underdoped.

Films were prepared by molecular-beam epitaxy (MBE) on 12.7 mm
$\times$ 12.7 mm $\times$ 0.35 mm SrTiO$_{3}$ substrates as
detailed elsewhere \cite{naito01,naito02,naito03}. The same
procedures and parameters were used for all films of a given
compound. Table I summarizes film properties. Ce concentrations
are measured by inductively coupled plasma spectroscopy and are
known to better than $\pm 0.005$. We refer to the LCCO film with
$x = 0.112$ and the PCCO film with $x = 0.145$ as ``optimally
doped'', although the optimal $x$ may be slightly smaller than
these values \cite{skinta02}. The optimal PCCO film is film P3
from Ref. 8. The films are highly \textit{c}-axis oriented, and
their \textit{ab}-plane resistivities, $\rho(T)$ in Fig. 1, are
low. Resistivities in our e-doped films are lower than in e-doped
crystals \cite{kokales02} and high-quality crystals of
La$_{2-x}$Sr$_x$CuO$_4$ \cite{ando}, the h-doped cousin of PCCO
and LCCO. $\rho(T)$ in our films decreases monotonically with
increasing doping, and $\rho(T)$ just above $T_c$ decreases by a
factor of two between under- and optimal doping.

We measure $\lambda^{-2}(T)$ with a low frequency two-coil mutual
inductance technique described in detail elsewhere
\cite{turneaure01}. A film is centered between two small coils,
and a current at about 50 kHz in one coil induces eddy currents in
the film. Currents are approximately uniform through the film
thickness. Data have been measured to be independent of frequency
for $10$ kHz $\leq f \leq 100$ kHz. Magnetic fields from the
primary coil and the film are measured as a voltage across the
secondary coil. We have checked that the typical excitation field
(100 $\mu$Tesla $\perp$ to film) is too small to create vortices
in the film. Because the coils are much smaller than the film, the
applied field is concentrated near the film's center and
demagnetizing effects at the film perimeter are irrelevant. All
data presented here are in the linear response regime.

The film's sheet conductivity, $\sigma (T)d = \sigma _{1} (T)d -
i\sigma_{2}(T)d$ with $d$ the film thickness, is deduced from the
measured mutual inductance. $\sigma_{1}$ is large enough to be
detectable only near $T_c$. We define $T_c$ and $\Delta T_c$ to be
the temperature and full-width of the peak in $\sigma_1$. $\lambda
^{-2}(T)$ is obtained from the imaginary part of the conductivity
as $\lambda ^{-2}(T) \equiv \mu _{0} \omega \sigma _{2}(T)$, where
$\mu _{0}$ is the magnetic permeability of vacuum. Experimental
noise is typically 0.2\% of $\lambda^{-2}(0)$ at low temperatures
and is at least partly due to slow drift in amplifier gain. The
10\% uncertainty in $d$ is the largest source of error in $\lambda
^{-2}(T)$. This uncertainty does not impact the temperature
dependence of $\lambda ^{-2}(T)/\lambda ^{-2}(0)$. As the films
were grown in the same MBE apparatus on successive runs, we
estimate the relative uncertainty in $\lambda^{-2}(0)$ in each
material to be $\pm 5$\%.

Measurement of $\sigma_1(T)$ is a stringent test of film quality,
as inhomogeneities in any layer will increase $\Delta T_c$. Figure
2 displays $\sigma_1(T)$ measured at 50 kHz for each film. $\Delta
T_c$ is typically $\leq 1$ K and indicates excellent film quality.
Structure in $\sigma_1$ is due to layers with slightly different
$T_c$'s. The small peak at 19.7 K in the data from the overdoped
LCCO film is probably due to a tiny bad spot at the film edge. No
corresponding feature in $\lambda^{-2}(T)$ is apparent at 19.7 K
(Fig. 3), indicating that this transition is unimportant to
analysis of the data. $T_c$'s determined from resistivity (Fig. 1)
and penetration depth measurements (Figs. 2 and 3) are identical.

A few more words about film quality are in order. Maximum $T_c$'s
of e-doped films \cite{alff01,skinta01,alff02} are the same as or
superior to those of e-doped crystals
\cite{armitage,kokales01,prozorov,kokales02}. Resistivities of
films are lower \cite{kokales02,fournier}. Crystals have intrinsic
homogeneity problems -- e.g., Ce-poor surfaces and gradients in Ce
content \cite{drews,skelton} -- as the reduction process required
to remove interstitial apical oxygen can cause phase decomposition
in bulk samples \cite{naito01}. Optimized LCCO crystals have yet
to be grown, so a comparison of LCCO films with crystals is
impossible \cite{naito03}.

Figure 3 displays $\lambda^{-2}(T)$ for all films. The most
important feature is that the evolution of $\lambda^{-2}(T)$ with
doping is the same in PCCO and LCCO, despite some quantitative
differences, such as the optimal values of $T_c$ and
$\lambda^{-2}(0)$ in LCCO being 20\% higher and 30\% lower,
respectively, than in PCCO. For a given compound,
$\lambda^{-2}(0)$ is about the same at optimal and overdoping, but
is a factor of 2 smaller at underdoping. Upward curvature in
$\lambda^{-2}(T)$ near $T_c$ appears only at and above optimal
doping. We emphasize that this feature \emph{is not} due to
inhomogeneity \cite{note01}. A more thorough analysis of the data
\cite{skinta02} reveals that the upward curvature can be ascribed
to the energy dependence of the density of states and is an
important aspect of e-doped film behavior.

We now examine the low-temperature behavior of $\lambda^{-2}(T)$,
shown in Figs. 4 and 5. We have previously found \cite{skinta01}
that $\lambda^{-2}(T)/\lambda^{-2}(0)$ in optimally doped PCCO
films is reproducibly exponential at low temperatures and obeys
the equation
\begin{equation}
\label{eq:02} \lambda^{-2}(T) \sim \lambda^{-2}(0) \left[ 1 -
C_{\infty} e^{-D/t} \right],
\end{equation}
where $t \equiv T/T_c$ and $D=0.85=\Delta_{min}/{k_BT_c}$. In
optimal LCCO (Fig. 5), an exponential fits $\lambda^{-2}(T)$ with
a best-fit value for the minimum gap of 0.73 $k_B T_c$. A
quadratic fit lies outside the experimental noise level and is
therefore unacceptable. In overdoped PCCO and LCCO (Fig. 4), the
first $\sim 5$\% drop in $\lambda^{-2}(T)/\lambda^{-2}(0)$ also
displays an exponential temperature dependence, with values of $D$
(0.55 and 0.46, respectively) substantially smaller than at
optimal doping. A best quadratic fit to overdoped PCCO data lies
outside the experimental noise in places and is statistically
poorer. For overdoped LCCO, the data are clearly very flat at low
temperatures, and a quadratic fit is extremely poor.

In underdoped PCCO and LCCO (Fig. 5, lower curves), the first
$\sim 5$\% drop in $\lambda^{-2}(T)/\lambda^{-2}(0)$ is consistent
with quadratic behavior. Lower experimental temperatures are
needed to rule out an exponential dependence with a very small gap
($D=0.46$ and 0.60 in underdoped PCCO and LCCO, respectively) in
these films. Values of $T_0$ from best fits of $1 - (T/T_0)^2$ to
$\lambda^{-2}(T)/\lambda^{-2}(0)$ are 20.3 K and 39.3 K for
underdoped PCCO and LCCO, respectively.

The picture that emerges from the data presented here, and from
data on many other films \cite{skinta02}, is that there is some
sort of transition near optimal doping. Three different features
of the superfluid density -- low-$T$ magnitude, near-$T_c$
curvature, and low-$T$ temperature dependence -- change abruptly.
In our PCCO films, the changes occur over a doping range with
essentially the same $T_c$. We note that there is an abrupt
transition in the behavior of h-doped cuprates near optimal
doping, which is associated with the onset of a pseudogap
\cite{timusk}. It may be coincidental that there are transitions
in e-doped and h-doped cuprate behavior near optimal doping.

We can only speculate as to the nature of the transition. On the
basis of the transition from quadratic to exponential behavior at
low temperatures, we surmise that the pairing symmetry changes
from \textit{d}-wave to \textit{s}-wave near optimal doping. This
conclusion is bolstered by recent tunnelling measurements
\cite{biswas} that are also consistent with a \textit{d}- to
\textit{s}-wave pairing transition near optimal doping in PCCO.
There is no \textit{d}-wave model that predicts flatter-than-$T^2$
behavior for the superfluid density at low temperatures
\cite{annett,hirschfeld,kosztin}. We note that the picture would
be clearer if the underdoped films exhibited a crossover from
quadratic to linear behavior at low temperatures, as predicted for
weakly disordered \textit{d}-wave superconductors
\cite{annett,hirschfeld,kosztin}.

We have presented high-precision measurements of $\lambda^{-2}(T)$
in PCCO and LCCO films at various dopings near optimal. Film
quality is demonstrably high. The two compounds, despite
quantitative differences in $T_c$ and other parameters, behave
similarly with doping. $\lambda^{-2}(0)$ increases rapidly as
optimal doping is approached from below and is roughly constant
above optimal doping. Upward curvature in $\lambda^{-2}(T)$ near
$T_c$, \emph{not} associated with film inhomogeneity, develops at
optimal doping and grows with overdoping. At low $T$, the
temperature dependence of $\lambda^{-2}(T)$ is exponential at
optimal and overdoping, but is quadratic at underdoping.
Exponential behavior is consistent with a gapped state, e.g.,
\textit{s}-wave superconductivity, while $T^2$ is usually
associated with \textit{d}-wave superconductivity. This apparent
transition in pairing symmetry would reconcile contradictory
literature results on e-doped cuprates, if nominally optimally
doped samples that exhibit \textit{d}-wave characteristics are in
reality underdoped.

\begin{figure}[htb]
\centerline{\epsfxsize=3.0in \epsfbox{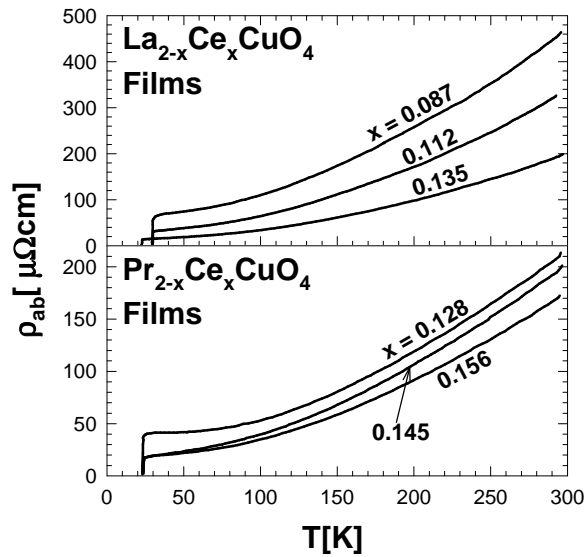}} \vspace{0.1in}
\caption{\textit{ab}-plane resistivities, $\rho(T)$, of six
electron-doped films. For resistivities just above $T_c$, see
Table I.} \vspace{-0.1in}
\end{figure}

\begin{figure}[htb]
\centerline{\epsfxsize=3.0in \epsfbox{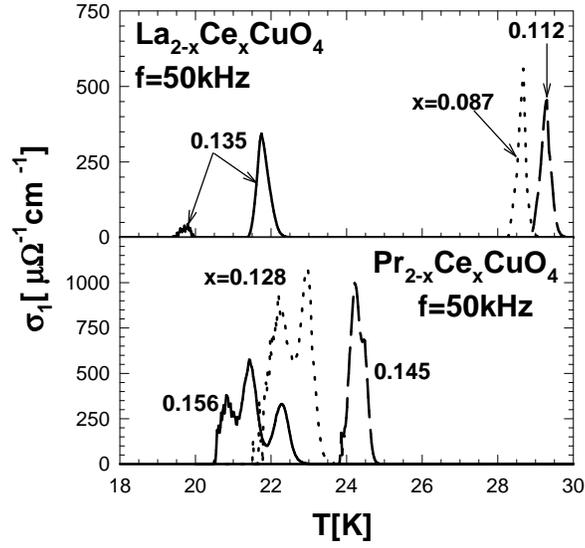}} \vspace{0.1in}
\caption{$\sigma_1(T)$ at 50 kHz in six electron-doped films.
$T_c$ and $\Delta T_c$ are temperature and full-width of peak in
$\sigma_1$.} \vspace{-0.1in}
\end{figure}

\begin{figure}[htb]
\centerline{\epsfxsize=3.0in \epsfbox{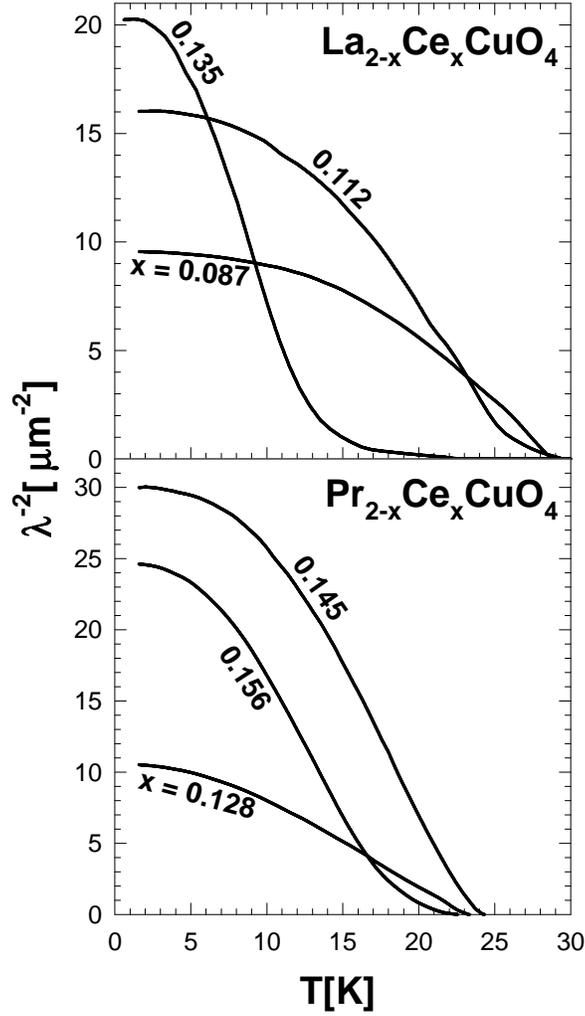}} \vspace{0.1in}
\caption{$\lambda^{-2}(T)$ for six electron-doped films. Relative
uncertainty in $\lambda^{-2}(0)$ in each material is $\sim 10$\%.}
\vspace{-0.1in}
\end{figure}

\begin{figure}[htb]
\centerline{\epsfxsize=3.0in \epsfbox{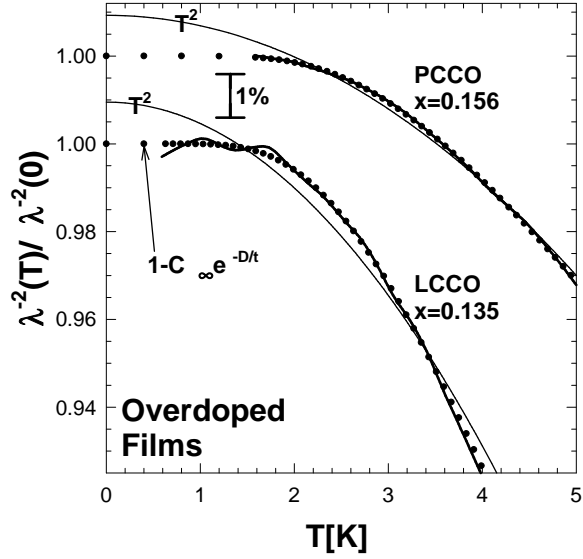}} \vspace{0.1in}
\caption{First $\sim 5$\% drop in
$\lambda^{-2}(T)/\lambda^{-2}(0)$ for overdoped PCCO and LCCO
films (thick lines), offset for clarity. Dotted curves are
exponential fits, $1 - C_{\infty} e^{-D/t}$ with $t \equiv T/T_c$,
to the data over this range. Thin solid lines are best quadratic
fits of the form $ 1 - (T/T_0)^2$. Exponential fits are visibly
superior.} \vspace{-0.1in}
\end{figure}

\begin{figure}[htb]
\centerline{\epsfxsize=3.0in \epsfbox{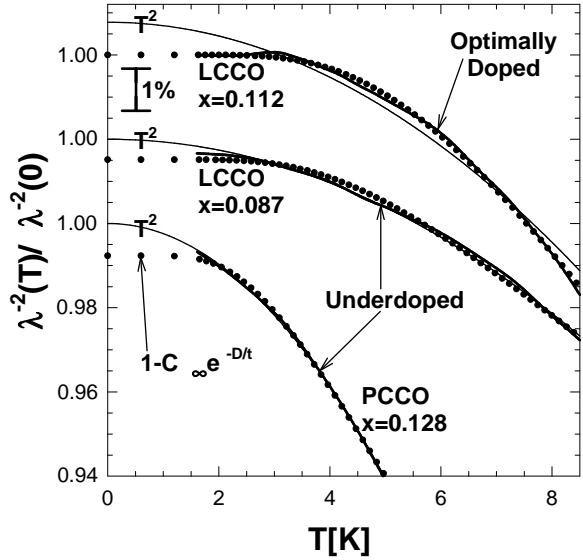}} \vspace{0.1in}
\caption{First $\sim 5$\% drop in
$\lambda^{-2}(T)/\lambda^{-2}(0)$ for optimally doped (upper
curve) and underdoped (lower curves) electron-doped films, offset
for clarity. Dotted curves are exponential fits, $1 - C_{\infty}
e^{-D/t}$ with $t \equiv T/T_c$, to the data over this range. Thin
solid lines are best quadratic fits of the form $ 1 - (T/T_0)^2$.
Optimally doped LCCO is exponential, and the underdoped films are
quadratic, at low temperatures.} \vspace{-0.1in}
\end{figure}

\pagebreak

\begin{table}
\caption{Properties of six electron-doped films. Ce doping, $x$,
is measured by inductively coupled plasma spectroscopy and is
known to better than $\pm 0.005$. $d$ is film thickness. $T_{c}$
and $\Delta T_{c}$ are location and full-width of peak in
$\sigma_{1}$. Absolute uncertainty in $\lambda^{-2}(0)$ is
$\pm$10\%. $\rho$($T_c + 5K$) is the \textit{ab}-plane resistivity
just above $T_c$. $C_{\infty}$ and $D$ are parameters of the
exponential fit in Eq. (1).} \label{table01}
\begin{tabular}{c c c c c c c c c}
Film & $x$ & $d$ [\AA] & $T_c$ [K] & $\Delta T_c$ [K] & $\lambda
(0)$ [\AA] & $\rho$($T_c + 5K$) [$\mu \Omega$ cm] & $C_{\infty}$ & $D$ \\
\hline \hline
underdoped LCCO & 0.087 & 1250 & 28.7 & 0.8 & 3200 & 67 & 0.32 & 0.60 \\
optimal LCCO    & 0.112 & 1250 & 29.3 & 0.9 & 2500 & 33 & 0.69 & 0.73 \\
overdoped LCCO  & 0.135 & 1250 & 21.7 & 1.0 & 2200 & 15 & 0.92 & 0.46 \\
\hline
underdoped PCCO & 0.128 & 1000 & 22.5 & 1.8 & 3100 & 40 & 0.41 & 0.46 \\
optimal PCCO    & 0.145 & 1000 & 24.2 & 1.0 & 1800 & 19 & 0.93 & 0.85 \\
overdoped PCCO  & 0.156 & 1000 & 21.5 & 2.4 & 2000 & 18 & 0.55 & 0.55 \\
\end{tabular}
\end{table}


\begin{references}
\bibitem{note01} We have observed upward curvature in
$\lambda^{-2}$ near $T_c$ before, but never at temperatures below
the fluctuation peak in $\sigma_1$. We ascribe upward curvature
with non-zero $\sigma_1$ to film inhomogeneity.
\end{references}
\end{document}